\documentclass[sigconf, natbib=false]{acmart}
\usepackage{graphicx}
\usepackage{array}
\usepackage{multirow}
\usepackage{booktabs}
\usepackage{enumitem}
\usepackage{subcaption}
\usepackage{amsmath}
\usepackage{hhline}

\AtBeginDocument{%
  }

\setcopyright{acmlicensed}
\copyrightyear{2018}
\acmYear{2018}
\acmDOI{XXXXXXX.XXXXXXX}
\acmConference[Conference acronym 'XX]{Make sure to enter the correct
  conference title from your rights confirmation email}{June 03--05,
  2018}{Woodstock, NY}
\acmISBN{978-1-4503-XXXX-X/2018/06}

\RequirePackage[
  datamodel=acmdatamodel,
  style=acmnumeric,
  ]{biblatex}

\usepackage{amsmath}
\begin{document}

\title{Revisiting scalable sequential recommendation with Multi-Embedding Approach and Mixture-of-Experts}

\author{Qiushi Pan}
\email{pqs@mail.ustc.edu.cn}
\affiliation{%
  \institution{University of Science and Technology of China}
  \city{Hefei}
  \state{Anhui}
  \country{China}
}

\author{Hao Wang}
\email{wanghao3@ustc.edu.cn}
\affiliation{%
  \institution{University of Science and Technology of China}
  \city{Hefei}
  \state{Anhui}
  \country{China}
}

\author{Guoyuan An}
\email{anguoyuan@h-partners.com}
\affiliation{%
  \institution{Huawei Noah’s Ark Lab}
  \city{Singapore}
  \state{Singapore}
  \country{Singapore}
}

\author{Luankang Zhang}
\email{zhanglk5@mail.ustc.edu.cn}
\affiliation{%
  \institution{University of Science and Technology of China}
  \city{Hefei}
  \state{Anhui}
  \country{China}
}

\author{Wei Guo}
\email{guowei67@huawei.com}
\affiliation{%
  \institution{Huawei Noah’s Ark Lab}
  \city{Singapore}
  \state{Singapore}
  \country{Singapore}
}
\author{Yong Liu}
\email{liu.yong6@huawei.com}
\affiliation{%
  \institution{Huawei Noah’s Ark Lab}
  \city{Singapore}
  \state{Singapore}
  \country{Singapore}
}



\begin{abstract}
  In recommendation systems, how to effectively scale up recommendation models has been an essential research topic. While significant progress has been made in developing advanced and scalable architectures for sequential recommendation(SR) models, there are still challenges due to items' multi-faceted characteristics and dynamic item relevance in the user context. To address these issues, we propose Fuxi-MME, a framework that integrates a multi-embedding strategy with a Mixture-of-Experts (MoE) architecture. Specifically, to efficiently capture diverse item characteristics in a decoupled manner, we decompose the conventional single embedding matrix into several lower-dimensional embedding matrices. Additionally, by substituting relevant parameters in the Fuxi Block with an MoE layer, our model achieves adaptive and specialized transformation of the enriched representations. Empirical results on public datasets show that our proposed framework outperforms several competitive baselines.
\end{abstract}


\received{20 February 2007}
\received[revised]{12 March 2009}
\received[accepted]{5 June 2009}

\maketitle
\section{Introduction}

Recommendation systems \cite{kang2018self, zhang2024wukong, he2014practical,liu2023user,wang2025generative,wang2025universal,wang2025mf,wang2019mcne,wang2021hypersorec,wang2021decoupled,tong2024mdap,zhang2024unified,xu2024multi} aim to analyze users’ historical behaviors to identify and present items that align with their potential interests. In industrial applications, Sequential Recommendation (SR) \cite{zhou2020s3, zhai2024actions,zhang2025td3,yin2024learning,yin2023apgl4sr,wang2024denoising,han2024efficient,han2023guesr} has become an indispensable component, as it focuses on predicting a user’s next action by modeling the temporal dynamics of their interaction history. This subfield has consequently attracted substantial attention from both academia and industry.

In recent years, the field has embraced the autoregressive paradigm \cite{kang2018self, deng2025onerec}, drawing inspiration from advances in natural language processing \cite{touvron2023llama, yang2025qwen3}. Early implementations, such as GRU4Rec \cite{hidasi2015session}, utilized Recurrent Neural Networks (RNNs) \cite{hochreiter1997long, medsker2001recurrent} to capture short-term user preferences but struggled with long-range patterns due to recursive state updates and gradient issues. Subsequently, SASRec \cite{kang2018self} marked a significant leap forward by incorporating a Transformer-based \cite{vaswani2017attention} architecture that leverages self-attention to capture more complex and dynamic user preferences. Despite these advancements, scalability remains a persistent challenge—simply enlarging model size does not guarantee improved performance \cite{shen2024optimizing}. To overcome this, contemporary research has turned to scaling laws \cite{kaplan2020scaling,guo2024scaling,shen2025optimizingsequentialrecommendationmodels}, revealing that model performance can grow predictably with increases in parameters, data, and compute. For instance, HSTU \cite{zhai2024actions} modified the attention mechanism for large and non-stationary vocabulary and proposes the M-FALCON algorithm to speed up model inference, which exhibits scaling-up effects in recommendation. Similarly, Fuxi-Alpha \cite{ye2025fuxi} presents a novel architecture that disentangles temporal, positional, and semantic features, offering a more scalable and modular design. Together, these works demonstrate that autoregressive frameworks can successfully harness scaling laws for considerable performance enhancements in recommendation.

Despite the considerable progress achieved in designing scalable architectures for SR, most existing studies emphasize model structure while \textbf{overlooking the embedding layer}—a component that often constitutes the majority of parameters and represents a major bottleneck in expressive capacity \cite{jiang2021xlightfm, wang2025universal}. Consequently, the ability to capture rich item features is often limited by the constraints of a single, monolithic embedding vector. While some approaches attempt to mitigate this by increasing embedding dimensionality, such brute-force scaling yields diminishing returns \cite{guo2023embedding}. Drawing inspiration from Large Language Models (LLMs) \cite{shen2024exploring,lv2025costeer,shen2025genkienhancingopendomainquestion,yin2024entropy,gu2025rapid} and their capacity to represent multifaceted semantic spaces \cite{zhang2024gme,huang2024chemeval}, we argue for a more principled rethinking of the embedding layer itself. By rearchitecting the embedding space, we can more effectively model complex item attributes and alleviate the representational bottleneck that limits current SR systems.

However, advancing this direction introduces new challenges. Research on embedding structures for sequential recommendation remains limited, and two critical challenges:
(1) An item’s identity is inherently multi-faceted—encompassing attributes such as category, brand, and style—yet a single dense embedding struggles to disentangle these heterogeneous signals, constraining the model’s expressiveness. \textbf{Designing a mechanism that can represent such diverse features in a decoupled and structured manner is therefore essential.}
(2) The relevance of item attributes varies with user context: one user may prioritize brand affinity, while another focuses on visual style. \textbf{Capturing these differences requires a dynamic, input-aware mechanism that can adaptively transform and weigh features according to sequence context.}

To tackle these challenges, we propose \textbf{Fuxi-MME}, a framework that integrates a multi-embedding strategy with a Mixture-of-Experts (MoE) architecture built upon the high-performing Fuxi-Alpha model. Our approach decomposes the conventional single embedding into multiple lower-dimensional sub-embeddings, each responsible for capturing a distinct facet of an item’s identity. This design enhances expressiveness without inflating parameter count and can be seamlessly integrated as a plug-and-play component. Furthermore, to enable adaptive and specialized processing of these enriched representations, we introduce an MoE layer within the Fuxi block. This layer serves as a dynamic routing mechanism, directing sequence inputs to specialized expert networks based on their contextual characteristics. Extensive experiments show that the synergy between expressive embeddings and adaptive experts enables end-to-end learning that effectively models heterogeneous user behaviors and nuanced feature interactions.
The main contributions of this paper are summarized as follows:
\begin{itemize}
\item We present Fuxi-MME, a framework that synergizes multi-embedding and expert routing to achieve more context-aware and adaptive sequence modeling.
\item We propose a multi-faceted embedding strategy for sequential recommendation, designed to overcome the representational bottleneck of conventional single-vector embeddings in a parameter-efficient manner.
\item We develop a MoE module tailored for Transformer-based architectures, enabling dynamic, input-dependent feature transformation and enhancing interaction modeling.
\item Comprehensive experiments on three public benchmark datasets demonstrate that Fuxi-MME establishes new state-of-the-art results. Further studies confirm the framework’s effectiveness.
\end{itemize}

\section{Related Work}

\subsection{Sequential Recommendation}
Sequential recommendation \cite{boka2024survey,pan2024survey,xu2019survey} aims to predict a user's next interaction based on their historical behavior. Early approaches predominantly relied on Markov chains \cite{he2016fusing, rendle2010factorizing} to model transition probabilities between items. With the rise of deep learning, subsequent research adopted neural architectures such as recurrent neural networks (RNNs) \cite{hidasi2018recurrent,hidasi2015session} and convolutional neural networks (CNNs) \cite{chen2022double,yan2019cosrec} to better capture complex temporal dependencies.

More recently, Transformer-based architectures have achieved state-of-the-art performance in this domain. For instance, SASRec \cite{kang2018self} formulates the task as an autoregressive prediction problem, employing self-attention to model user behavior sequences. In contrast, BERT4Rec \cite{sun2019bert4rec} utilizes a bidirectional self-attention mechanism with a masked item prediction objective, allowing the model to leverage both past and future context for richer item representations. 

Beyond attention-based models, generative paradigms such as GANs \cite{chen2024collaborative,wang2021siamese}, VAEs \cite{zhou2023vcgan}, and diffusion models \cite{wu2024diffusion,xie2024breaking} have been explored to improve sequence modeling and uncertainty estimation. Furthermore, large language models (LLMs) \cite{wu2024survey} have recently been incorporated into recommendation systems to generate semantic input embeddings \cite{li2023exploring,li2023ctrl,shen2023towards,shen2024exploring}, thereby introducing external knowledge and improving representation quality.

\subsection{Scaling Recommendation Models}
Building on the generative perspective, recent studies \cite{deldjoo2024recommendation,hou2025generative,liu2024multi,wang2025generative,ye2025fuxi2} have reformulated recommendation as a sequence generation problem, where the model autoregressively produces tokens corresponding to item identifiers. A common research trend has been to scale model parameters to enhance representational and generative capacity \cite{guo2024scaling}.

Following this direction, TIGER \cite{rajput2023recommender} employs RQ-VAE \cite{lee2022autoregressive} to construct semantically meaningful discrete item codes and trains a Transformer to predict them in sequence. HSTU \cite{zhai2024actions} extends this idea by scaling to trillion-parameter models that unify heterogeneous user behaviors into a single generative sequence, demonstrating consistent performance gains with model size. Wukong \cite{zhang2024wukong} introduces a stackable layer design for scalable architecture expansion, while another line of work \cite{guo2023embedding} mitigates embedding collapse using a multi-embedding strategy.

This generative paradigm has recently been extended across the entire recommendation pipeline \cite{guo2025onesug,zhang2025killing}. For example, OneRec \cite{deng2025onerec} replaces traditional ID-based item representations with semantic encodings, integrates Delayed Propagation Optimization (DPO) \cite{rafailov2023direct} into a Mixture-of-Experts (MoE) \cite{dai2024deepseekmoe,zoph2022designing} Transformer framework, and unifies multi-stage learning into a single end-to-end training process. Although these methods achieve strong performance and scalability, they primarily focus on architectural scaling, while the potential of scaling input representations remains underexplored.

\section{Problem Definition}

Let $\mathcal{U} = \{u_1, u_2, ..., u_{|\mathcal{U}|}\}$ represent the set of all users and $\mathcal{I} = \{i_1, i_2, ..., i_{|\mathcal{I}|}\}$ represent the set of all unique items in the system. The historical interactions of each user are captured as a chronologically ordered sequence. For a given user $u \in \mathcal{U}$, their interaction history is denoted as $S^u = (i_1^u, i_2^u, ..., i_{n_u}^u)$, where $i_t^u \in \mathcal{I}$ is the item that user $u$ interacted with at time step $t$, and $n_u$ is the total length of the sequence for that user.

For practical implementation and efficient batch processing in deep learning models, these variable-length sequences are typically converted to a fixed length, $n$. If a user's sequence is shorter than $n$ ($n_u < n$), the sequence is padded with a special $\langle \text{PAD} \rangle$ token at the beginning. If the sequence is longer than $n$ ($n_u > n$), only the most recent $n$ interactions are retained, as these are generally the most indicative of the user's current interests.

The fundamental task of sequential recommendation is to predict the next item, $i_{n+1}^u$, that a user $u$ is most likely to interact with, given their historical interaction sequence $S^u = (i_1^u, ..., i_n^u)$. Formally, this is a probabilistic task of finding the item that maximizes the conditional probability:

\begin{equation}
    \hat{i}_{n+1}^u = \underset{i \in \mathcal{I}}{\arg\max} \, P(i_{n+1}=i|S^u)
\end{equation} 

A sequential recommendation model, parameterized by $\theta$, is trained to learn this probability distribution $P( \cdot | S^u; \theta)$. This is typically achieved by designing a model $f_\theta$ that performs two main functions:
\begin{enumerate}
    \item \textbf{Sequence Encoding:} It encodes the input sequence $S^u$ into a high-dimensional vector representation, $h^u_n \in \mathbb{R}^d$, which is intended to capture the user's current preferences and intent.
    \item \textbf{Item Scoring:} It computes a relevance score between the user's representation $h^u_n$ and the embedding vector $e_i \in \mathbb{R}^d$ of every candidate item $i \in \mathcal{I}$. The probability is then often estimated via a softmax function over these scores.
\end{enumerate}

In practice, the goal is not just to predict a single item but to generate a ranked list of the Top-K most probable items for recommendation.

\section{Methodology}

In this section, we introduce the proposed \textbf{Fuxi-MME} framework. We begin by providing a concise overview of the Fuxi-$\alpha$ model, which serves as the foundational architecture for our work. Subsequently, we detail our two primary contributions: the \textbf{Multi-embedding Approach} designed to capture the multifaceted nature of items, and the \textbf{Mixture-of-Experts (MoE) enhanced Fuxi Block}, which enables dynamic and specialized feature processing. Finally, we describe the model's training objective.

\subsection{The Foundational Backbone Fuxi-$\alpha$}
We select Fuxi-$\alpha$ \cite{ye2025fuxi} as our base model due to its demonstrated state-of-the-art performance and its scalable architecture, which is specifically designed to handle large-scale sequential recommendation tasks. The core of Fuxi-$\alpha$ is a stack of $L$ identical decoder layers, referred to as \textit{Fuxi blocks}. Each Fuxi block contains two main sub-layers: an Adaptive Multi-channel Self-attention (AMS) layer and a Multi-stage Feed-Forward Network (MFFN).

\begin{figure}[t]
\centering
\includegraphics[width=0.95\linewidth]{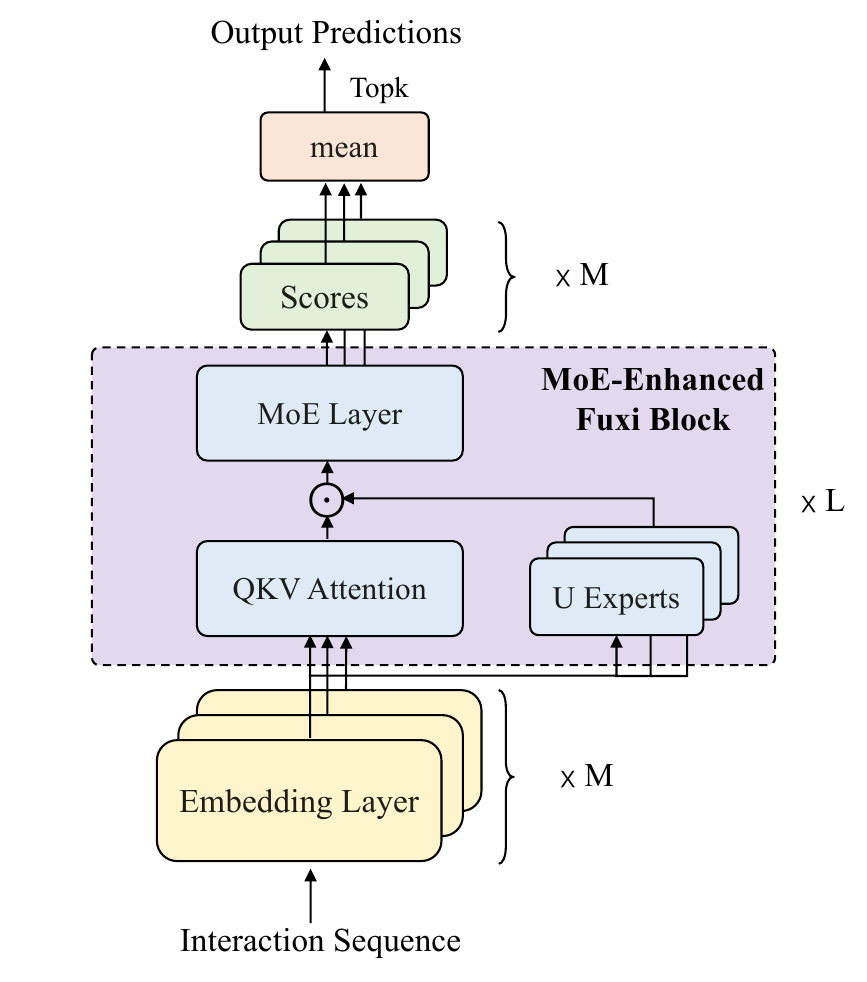}
\caption{Overview of framework.} \label{overview}
\end{figure}

\subsubsection{Adaptive Multi-channel Self-attention (AMS)}
Unlike standard self-attention, the AMS layer models user sequences through three distinct channels to capture different aspects of item-to-item relationships:
\begin{enumerate}
    \item \textbf{Semantic Channel:} This channel captures the inherent feature-based relationships between items, akin to the standard self-attention mechanism in Transformers. For an input sequence representation $x^{l-1}$ from the previous layer, the computation is as follows:
    \begin{equation}
        \hat{x}^l = \text{RMSN}(x^{l-1})
    \end{equation}
    \begin{equation}
        q^l = \phi(\hat{x}^lW_q^l), \quad k^l = \phi(\hat{x}^lW_k^l), \quad v^l = \phi(\hat{x}^lW_v^l)
    \end{equation}
    \begin{equation}
        a_h^l = \frac{1}{\sqrt{d_k}}\phi(q^l(k^l)^T)
    \end{equation}
    where $W_q^l, W_k^l, W_v^l \in \mathbb{R}^{d\times d_h}$ are learnable projection matrices for the attention mechanism. $\text{RMSN}$ denotes Root Mean Square Layer Normalization, and $\phi$ is a non-linear activation function, typically SiLU.

    \item \textbf{Temporal and Positional Channels:} To explicitly model the sequence order and time intervals, Fuxi-$\alpha$ incorporates two additional channels whose attention scores are calculated separately:
    \begin{equation}
        (a_t^l)_{i,j} = \alpha(t_j-t_i), \quad (a_p^l)_{i,j} = \beta_{j-i}
    \end{equation}
    Here, $(a_t^l)_{i,j}$ represents the temporal bias based on the timestamp difference between items $i$ and $j$, while $(a_p^l)_{i,j}$ represents the relative positional bias. The parameters $\alpha$ and $\beta$ are learnable mappings that encode these biases.
\end{enumerate}
The outputs from these three channels are then aggregated and combined with a gated projection, similar to the mechanism in HSTU \cite{zhai2024actions}:
\begin{equation}
    h^l = \text{RMSN}(\text{concat}(a_h^lv^l, a_t^lv^l, a_p^lv^l) \otimes \phi(x^{l-1}W_u^l))
\end{equation}
where $\otimes$ denotes element-wise multiplication, which acts as a dynamic gate to control the information flow from the attention module.
\subsubsection{Multi-stage Feed-Forward Network (MFFN)}
The MFFN processes the output from the AMS layer through a two-stage process. First, a linear projection is applied, followed by a residual connection:
\begin{equation}
    o^l = h^lW_o^l + x^{l-1}
\end{equation}
Second, this intermediate representation is passed through a well-designed FFN structure that uses gating to control information flow and model complex feature interactions:
\begin{equation}
    x^l = \text{FFN}_l(\text{RMSN}(o^l)) + o^l
\end{equation}
\begin{equation}
    \label{eq:ffn}
    \text{FFN}_l(x) = (\phi(xW_1^l) \otimes (xW_2^l))W_3^l
\end{equation}
where $W_1^l, W_2^l \in \mathbb{R}^{d\times d_{FFN}}$ and $W_3^l \in \mathbb{R}^{d_{FFN}\times d}$ are learnable weight matrices. This gated structure has been shown to be more effective than standard FFNs.

\subsection{Multi-embedding Approach}
A primary limitation of existing sequential models is their reliance on a single, monolithic embedding vector for each item. This design forces a single vector to encode all of an item's multifaceted characteristics (e.g., category, brand, style, price), creating a representational bottleneck and potentially leading to embedding collapse \cite{guo2023embedding}, where distinct features become entangled.

Inspired by \cite{guo2023embedding}, we propose a multi-embedding approach to address this. Instead of a single embedding matrix, we introduce $M$ independent embedding matrices, $\{E_1, E_2, ..., E_M\}$, where each $E_i \in \mathbb{R}^{|I| \times (d/M)}$ and $|I|$ is the total number of items. This partitions the total embedding dimension $d$ into $M$ smaller, specialized "sub-spaces," with the crucial benefit of maintaining the same total parameter count as a standard embedding layer.

Given a user's interaction sequence $S_u=(i_1, i_2, ..., i_n)$, we map each item $i_t$ to $M$ corresponding sub-embeddings. This results in $M$ distinct input representation matrices $\{X_1^0, X_2^0, ..., X_M^0\}$, where $X_i^0 \in \mathbb{R}^{n \times (d/M)}$.

Critically, and in contrast to prior work that might process each representation stream independently, we feed all $M$ representation matrices into a single, shared sequential decoder (i.e., the stacked Fuxi blocks):
\begin{equation}
    \{X_1^L, ..., X_M^L\} = \text{Decoder}(\{X_1^0, ..., X_M^0\}; \Theta)
\end{equation}
where $\Theta$ represents the shared parameters of the Fuxi-$\alpha$ decoder. This design choice is deliberate: it allows the model to learn universal sequential patterns (e.g., temporal dynamics, general user behavior) across all embedding spaces through the shared parameters of $\Theta$, while still allowing each of the $M$ streams to preserve its specialized, disentangled feature information.

After processing through all $L$ decoder layers, we obtain $M$ final output representations. During prediction, these are used to compute scores for candidate items, which are then aggregated via mean pooling to produce the final recommendation score.


\subsection{MoE-enhanced Fuxi Block}
While the multi-embedding approach enriches the input representations, the model must also be able to process these diverse and multifaceted features adaptively. A standard, dense network applies the same transformations to all inputs, which is suboptimal when dealing with varied signals. To address this, we enhance the Fuxi block by incorporating a Mixture-of-Experts (MoE) architecture.

The MoE paradigm decouples model capacity from computational cost by activating only a sparse subset of the network for any given input. We integrate MoE layers to replace the dense FFN and the gating projection matrix $U$ within each Fuxi block. This allows the model to learn specialized "expert" networks and dynamically route different parts of the input sequence to the most appropriate experts.

Specifically, we replace the dense FFN from Equation \ref{eq:ffn} with a Sparse-Gated MoE layer. For a given input $x$, a lightweight gating network $G(x)$ calculates routing weights to select the top-$k$ experts:
\begin{equation}
    \text{MoE-Output}(x) = \sum_{i=1}^N G(x)_i \cdot \text{Expert}_i(x)
\end{equation}
where $N$ is the total number of experts (e.g., individual FFNs) and $G(x)_i$ is the weight assigned to the $i$-th expert for input $x$. The gating mechanism is designed to be sparse:
\begin{equation}
    G(x) = \text{Softmax}(\text{KeepTopK}(H(x), k))
\end{equation}
where $k$ is hyperparameter, ensuring only a small subset of experts are activated. The routing logits $H(x)$ are computed as:
\begin{equation}
    H(x) = x \cdot W_g + \text{StandardNormal}() \cdot \text{Softplus}(x \cdot W_{noise})
\end{equation}
Here, $W_g$ and $W_{noise}$ are learnable parameters of the gating network. The noise term is added during training to improve load balancing across experts, preventing a few experts from being consistently chosen. The $\text{KeepTopK}$ function sets the logits of all non-top-$k$ experts to $-\infty$, effectively zeroing out their weights after the Softmax operation:
\begin{equation}
    \text{KeepTopK}(v,k)_i = \begin{cases} v_i & \text{if } v_i \text{ is in the top } k \text{ elements of } v \\ -\infty & \text{otherwise} \end{cases}
\end{equation}
By integrating this MoE structure, Fuxi-MME can dynamically allocate its parameters based on the input, allowing for a significant increase in model capacity while keeping the computational cost per forward pass nearly constant. This is particularly effective for modeling the diverse patterns emerging from our multi-embedding representations.

\begin{table*}[h!]
\caption{Dataset statistics.}
\label{tab:dataset}
\renewcommand{\arraystretch}{1.1}
\begin{tabular}{llllll}
\hline
\textbf{Dataset} & \textbf{\#Users} & \textbf{\#Items} & \textbf{\#Interactions} & \textbf{Avg.Length} & \textbf{Sparsity} \\ \hline
Amazon-Books     & 694897           & 695761           & 10053086                & 14.47               & 99.99\%           \\
Amazon-Beauty    & 40226            & 57288            & 353962                  & 8.80                & 99.85\%           \\
Yelp             & 30431            & 20033            & 255492                  & 8.40                & 99.96\%           \\ \hline
\end{tabular}
\end{table*}

\subsection{Model Training and Optimization Objective}
\label{sec:training}

We train the Fuxi-MME framework in an end-to-end fashion using an autoregressive next-item prediction task. For a given user sequence $(i_1, i_2, ..., i_{n-1})$, the model's objective is to accurately predict the next item, $i_n$.

The final similarity score between the user's sequence representation at time $t$ and a candidate item $i$ is calculated by aggregating the scores from all $M$ embedding spaces:
\begin{equation}
    r(t,i) = \frac{1}{M}\sum_{k=1}^M \text{cosine\_similarity}(x_{t,k}^L, e_{i,k})
\end{equation}
where $x_{t,k}^L$ is the final output representation for the $t$-th item in the sequence from the $k$-th embedding space, and $e_{i,k}$ is the corresponding $k$-th sub-embedding of the candidate item $i$.

Due to the massive size of the item vocabulary in real-world scenarios, computing a full softmax over all items is computationally prohibitive. Therefore, we employ a sampled-softmax loss \cite{klenitskiy2023turning} for efficient training. For each positive instance (the true next item), we randomly sample a set of $N$ negative items from the item catalog. The model is then trained to maximize the score of the positive item relative to the negative ones.

The overall training loss is the negative log-likelihood over all sequences in the training set:
\begin{equation}
    \mathcal{L}_{\text{rec}} = -\sum_{u \in U}\sum_{t=1}^{n_u} \log \frac{\exp(r(t,i_t))}{\exp(r(t,i_t)) + \sum_{j \in I_{\text{neg}}} \exp(r(t,j))}
\end{equation}
where $I_{\text{neg}}$ is the set of $N$ randomly sampled negative items. This objective function effectively trains all components of the Fuxi-MME model, including the multiple embedding matrices, the shared Fuxi blocks, and the MoE layers, to work in concert for sequential recommendation.

\section{Experiments}

In this section, we present a comprehensive empirical evaluation of our proposed \textbf{Fuxi-MME} model. Our goal is to answer the following research questions:
\begin{itemize}
    \item \textbf{(RQ1)} Does Fuxi-MME outperform state-of-the-art sequential recommendation baselines across different datasets?
    \item \textbf{(RQ2)} What are the individual contributions of the multi-embedding approach and the Mixture-of-Experts (MoE) architecture to the model's overall performance?
    \item \textbf{(RQ3)} How sensitive is Fuxi-MME to its key hyperparameters, namely the number of embeddings ($M$) and the number of experts ($n$)?
    \item \textbf{(RQ4)} Is the placement of the MoE layer within the attention block optimal, and what does this imply about its function?
\end{itemize}
\subsection{Experimental Setup}
\subsubsection{Datasets.}
To evaluate the efficacy of our proposed method, we conduct extensive experiments on three public datasets. These datasets are derived from real-world online user interactions and are commonly adopted in sequential recommendation research. The statistics of these datasets are shown in \ref{tab:dataset}. Detailed descriptions are provided as follows:
\begin{itemize}
    \item \textit{Amazon-Books} and \textit{Amazon-Beauty}. The Amazon-Books and Amazon-Beauty datasets are subsets of the Amazon-Reviews dataset\footnote{http://jmcauley.ucsd.edu/data/amazon/}. It comprises real-world user interactions collected from the Amazon website, including both product metadata and user review records. The product metadata encompasses fields such as product ID, title, and category classifications. The user review records contain information such as user IDs, product IDs, review texts, ratings, and timestamps. 
    \item \textit{Yelp}\footnote{https://www.yelp.com/dataset}. The Yelp dataset is collected from the popular business website Yelp, comprises information about businesses, users, and reviews. In this work, we obtain a subset of the full dataset, following the preprocessing strategy in \cite{xie2022contrastive,chen2022intent,yin2024dataset}.
\end{itemize}

These three datasets are specifically chosen to provide a comprehensive testbed for our model. The two Amazon datasets represent dense e-commerce scenarios with distinct product feature distributions, while the Yelp dataset is known for its greater sparsity and different user behavior patterns related to local businesses. Success across these varied domains demonstrates the robustness and generalizability of a given approach.

\begin{table*}[t] 
 \centering 
 \caption{Overall Performance across three datasets. For each dataset, the best result is bolded while the second-best result is underlined.(p-value<0.05)} 
 \label{tab:overall_performance} 
 \renewcommand{\arraystretch}{1.3} 
 \setlength{\tabcolsep}{3.5pt} 
 \begin{tabular}{l|llll|llll|llll} 
 \hhline{=|====|====|====} 
 \multirow{2}{*}{\textbf{Model}} & \multicolumn{4}{c|}{\textbf{Amazon-books}} & \multicolumn{4}{c|}{\textbf{Amazon-beauty}} & \multicolumn{4}{c}{\textbf{Yelp}} \\ 
 \cline{2-13} 
 & \textbf{NG@10} & \textbf{NG@50} & \textbf{HR@10} & \textbf{HR@50} & \textbf{NG@10} & \textbf{NG@50} & \textbf{HR@10} & \textbf{HR@50} & \textbf{NG@10} & \textbf{NG@50} & \textbf{HR@10} & \textbf{HR@50} \\ 
 \hhline{=|====|====|====} 
 NARM &0.0109 &0.0188 &0.0209 &0.0580 &0.0201 &0.0305 &0.0361 &0.0840 &0.0172 &0.0314 &0.0338 &0.1003 \\ 
 SASRec &0.0202 &0.0324 &0.0378 &0.0944 &0.0313 &0.0445 &0.0553 &0.1160 &0.0186 &0.0333 &0.0363 &0.1052 \\ 
 GRU4Rec &0.0188 &0.0302 &0.0352 &0.0877 &0.0308 &0.0442 &0.0548 &0.1165 &0.0202 &\underline{0.0365} &0.0394 &\underline{0.1158} \\ 
 Tim4Rec & 0.0226&0.0358 &0.0421 &0.1030 &0.0291 &0.0421 &0.0520 &0.1120 &0.0173 &0.0328 &0.0347 &0.1072 \\ 
 Mamba4Rec &0.0221 &0.0344 &0.0405 &0.0970 &0.0290 &0.0418 &0.0508 &0.1097 &0.0173 &0.0322 &0.0344 &0.1039 \\ 
 HSTU &0.0255 & 0.0391&0.0470 &0.1096 &\underline{0.0337} &\underline{0.0485} &\underline{0.0587} &\underline{0.1270} &\underline{0.0205} &0.0356 &\underline{0.0399} &0.1102 \\ 
 Fuxi-alpha &\underline{0.0279} &\underline{0.0426} &\underline{0.0509} &\underline{0.1191} &0.0320 &0.0459 &0.0561 &0.1198 & 0.0195&0.0352 &0.0378 &0.1109 \\ 
 \hhline{-|----|----|----} 
 Fuxi-MME &\textbf{0.0294} &\textbf{0.0441} &\textbf{0.0534} &\textbf{0.1213} &\textbf{0.0358} &\textbf{0.0508} &\textbf{0.0626} &\textbf{0.1312} &\textbf{0.0225} &\textbf{0.0396} &\textbf{0.0443} &\textbf{0.1242} \\ 
 \hhline{=|====|====|====} 
 \end{tabular} 
 \end{table*}

\subsubsection{Compared Methods.}
We compare our proposed method with several classic and state-of-the-art baselines for sequential recommendation. This includes classic RNN-based models that set early benchmarks, modern Transformer-based architectures that represent the current state-of-the-art in modeling complex dependencies, and emerging State-Space Models (SSMs) which offer an efficient alternative for sequence modeling. This diverse set ensures a thorough and challenging comparison. The specific baselines are as follows:
\begin{itemize}
    \item GRU4Rec\cite{hidasi2015session}. GRU4Rec applies GRU layers to capture user preferences within interaction sequences.
    \item NARM\cite{li2017neural}. NARM combines RNN and attention mechanism to consider both user sequential behaviors and main purpose in sessions.
    \item SASRec\cite{kang2018self}. SASRec proposes a self-attention based sequential model to capture long-term semantics in users' interaction sequences.
    \item Mamba4Rec\cite{liu2024mamba4rec}. Mamba4Rec is the first work to utilize state space models for efficient sequential recommendation.
    \item Tim4Rec\cite{fan2025tim4rec}. Tim4Rec proposes a time-aware Mamba for sequential recommendation.
    \item HSTU\cite{zhai2024actions}. HSTU designs a new self-attention based sequential recommendation architecture for higher efficiency and scaling up effects.
    \item Fuxi-$\alpha$\cite{ye2025fuxi}. Fuxi-$\alpha$ introduces an Adaptive Multi-channel Self-attention mechanism to distinctly model temporal, positional and semantic features.
\end{itemize}
\subsubsection{Evaluation Protocols.}
We employ a leave-one-out strategy for evaluation: the last item is for testing, the second-to-last is for validation, and all others are for training. The partitioning of the dataset is consistent with \cite{zhai2024actions,ye2025fuxi}.
We use two widely adopted metrics for Top-$K$ recommendation lists: Hit Rate (HR@K) and Normalized Discounted Cumulative Gain (NDCG@K), with $K \in \{10, 50\}$. Crucially, we rank the ground-truth item against the entire item pool to ensure a fair and unbiased evaluation, avoiding biases from negative sampling during evaluation.



\subsubsection{Implementation Details}
To ensure fair evaluation of model performance, all models are implemented using PyTorch and trained on GPUs. We use the Adam \cite{kingma2014adam} optimizer with a learning rate of $1e^{-3}$. The sampled-softmax loss used a batch size of 256 with 128 negative samples per positive instance. We train all models for 100 epochs and use an early stop strategy, which is quitting training without improving the metrics for 10 epochs. For self-attention-based models, the number of layers is fixed at 2. The embedding dimension of baselines is searched in $[128,256,512]$ to compare with the multi-embedding approach. The embedding dimension of our proposed method is set to 128, and the number of embeddings is varied in $[1,2,4]$.

\subsection{Overall Performance (RQ1)}

The main results of our comparative evaluation are presented in Table \ref{tab:overall_performance}. The experimental findings clearly and consistently demonstrate the effectiveness of our proposed \textbf{Fuxi-MME} framework across all three benchmark datasets. We provide a detailed analysis below:

\begin{itemize}
    \item The results confirm the strength of Transformer-based models in sequential recommendation. Self-attention models  generally outperform RNN-based methods and the State-Space Models. This performance gap underscores the self-attention mechanism's superior ability to capture complex, long-range dependencies and model dynamic user preferences. While SSM-based models offer impressive efficiency, these results suggest that on standard benchmarks, their recommendation performance has not yet surpassed that of state-of-the-art Transformer architectures.

    \item Our proposed model, Fuxi-MME, sets a new state-of-the-art by a significant margin across all datasets and evaluation metrics. It consistently surpasses even the strongest and most recent baselines, including its own backbone, Fuxi-$\alpha$. For instance, on the challenging and sparse Amazon-Books dataset, Fuxi-MME improves upon the best baseline (Fuxi-$\alpha$) by 5.37\% in NDCG@10 and 4.91\% in HR@10. This substantial and consistent performance gain provides strong empirical evidence for our central thesis. The improvement is attributed to our novel architectural design: the multi-embedding approach provides a richer, more expressive representation of items by disentangling their multifaceted characteristics, and the MoE layer enables the model to process these nuanced features in an adaptive, input-dependent manner.
    \item Furthermore, comparing HSTU and Fuxi-$\alpha$, Fuxi-$\alpha$ performs better on the Amazon-Books dataset, while HSTU performs better on the other two datasets. This is because the embedding size is searched within the range of [128, 256, 512] when evaluating the two models, while for Fuxi-$\alpha$, its performance tends to decline when the embedding size increases to 512. This result empirically validates the core motivation of our paper. It suggests that a single, high-dimensional vector struggles to effectively organize diverse item features, leading to a representational bottleneck. Fuxi-MME addresses this by structuring the embedding space, leading to more consistent and scalable performance gains.
\end{itemize}

\subsection{Ablation Study (RQ2)}
\begin{table}[t]
\centering
\caption{Ablation Study.}
\label{tab:performance}
\renewcommand{\arraystretch}{1.3}
\resizebox{\columnwidth}{!}{%
\begin{tabular}{l|ccccc}
\hhline{=|=|====} 
\textbf{Model} & \textbf{NG@10} & \textbf{NG@50} & \textbf{HR@10} & \textbf{HR@50} & \textbf{MRR} \\
\hhline{-|-----} 
(a) w/o multi-embedding & 0.0306 & 0.0444 & 0.0543 & 0.1178 & 0.0277 \\
(b) ensemble & 0.0341 & 0.0489 & 0.0588 & 0.1267 & 0.0311 \\
(c) w/o MoE & 0.0344 & 0.0489 & 0.0598 & 0.1262 & 0.0311 \\
\hline
\textbf{Fuxi-MME} & \textbf{0.0358} & \textbf{0.0508} & \textbf{0.0626} & \textbf{0.1312} & \textbf{0.0322} \\
\hhline{=|=|====} 
\end{tabular}
}
\end{table}
To rigorously evaluate the individual contributions of our two primary architectural innovations, i.e. themulti-embedding approach and the Mixture-of-Experts (MoE) layer—we conduct a comprehensive ablation study. We design three distinct variants of our model and tested them on the Amazon-Beauty dataset. The results, presented in Table \ref{tab:performance}, allow us to systematically dissect the sources of Fuxi-MME's performance gains.

\begin{enumerate}[label=(\alph*)]
    \item \textbf{\emph{\textit{w/o multi-embedding}}:} This variant reverts to a standard, monolithic embedding layer with the same total dimension but removes the multi-embedding structure. It isolates the impact of our core representational hypothesis.
    \item \textbf{\emph{\textit{ensemble}}:} This variant trains $M$ separate Fuxi-$\alpha$ models, each with its own embedding space and its own decoder. The final predictions are generated by averaging their output scores. This tests whether Fuxi-MME's performance is a naive ensemble effect.
    \item \textbf{\emph{\textit{w/o MoE}}:} This variant retains the multi-embedding input but replaces the MoE-enhanced Fuxi blocks with the standard, dense Fuxi blocks. This quantifies the specific benefit of the adaptive, expert-based processing.
\end{enumerate}

\begin{figure}[t]
    \centering
    \begin{subfigure}{0.236\textwidth}
        \centering
        \includegraphics[width=\linewidth]{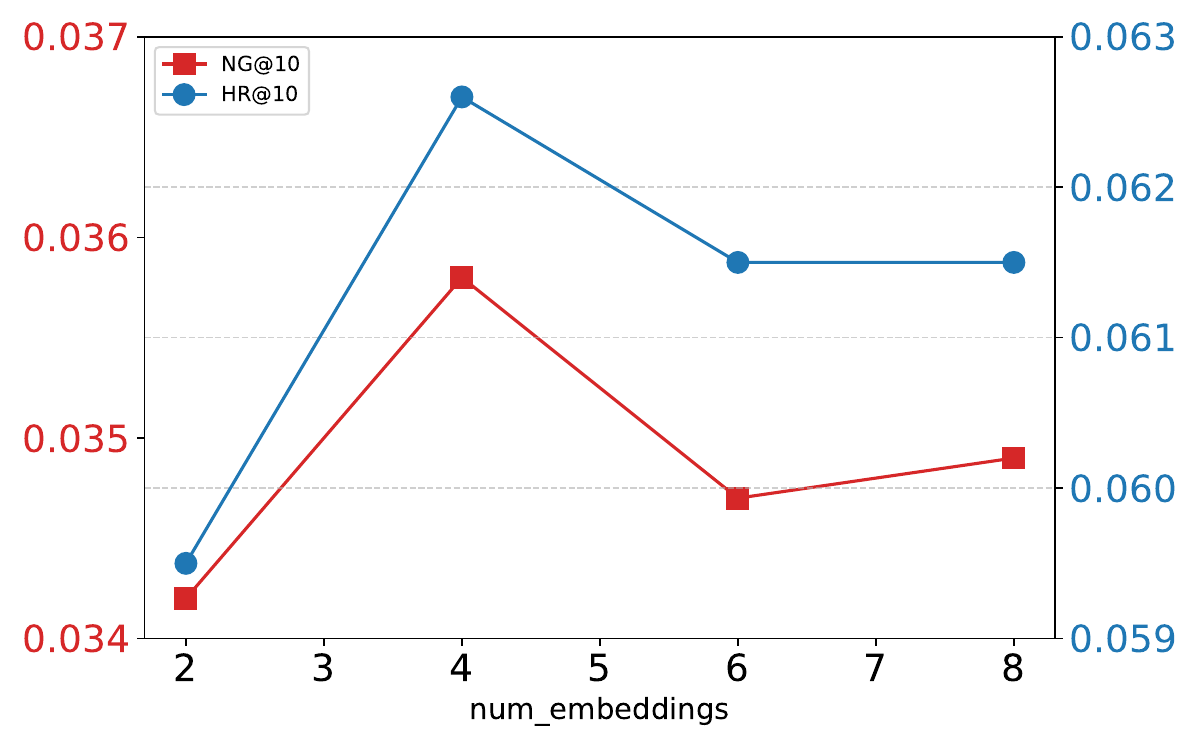}
        \caption{Impact of num embeddings.}
        \label{fig:subfig1numsample}
    \end{subfigure}%
    \hfill
    \begin{subfigure}{0.236\textwidth}
        \centering
        \includegraphics[width=\linewidth]{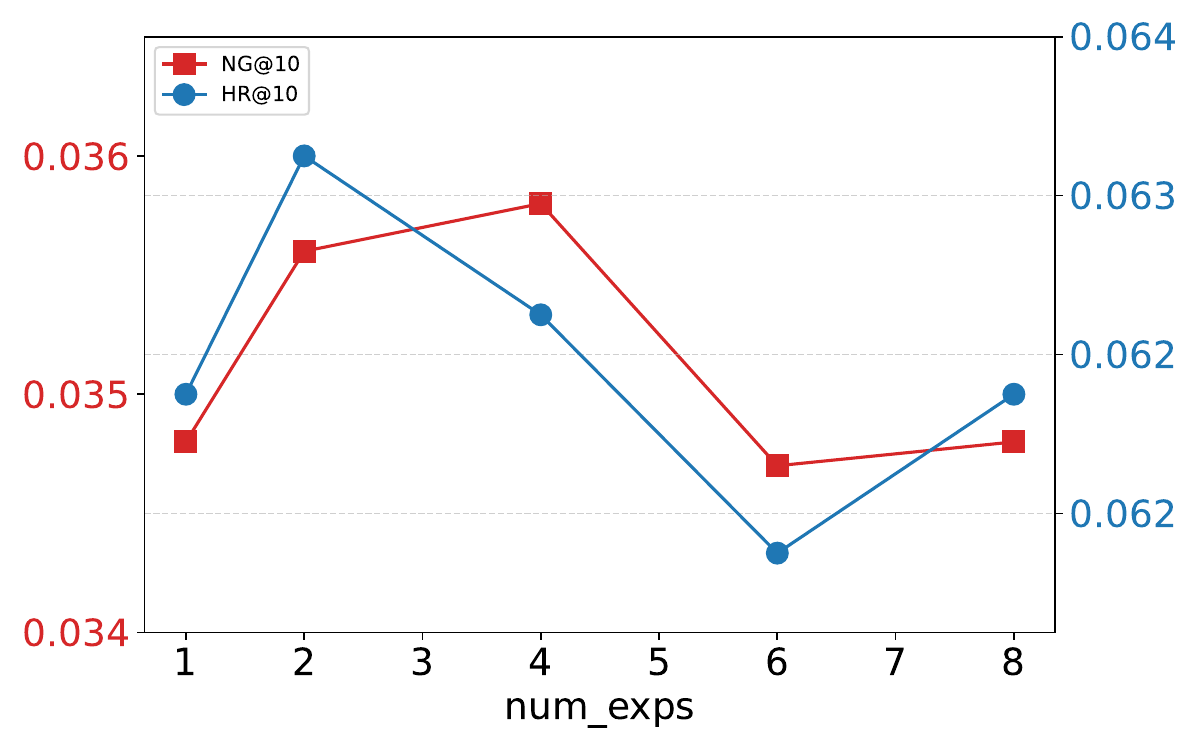}
        \caption{Impact of num exps.}
        \label{fig:subfig2numsample}
    \end{subfigure}
    \caption{Analysis of Hyper-parameter.}
    \label{fig:Hyper-parameter}
\end{figure}

Our analysis of the experimental results in Table  \ref{tab:performance} leads to the following key insights:
\begin{itemize}
    \item Comparing with (a), our proposed model achieves a significant improvement, highlighting the advantage of the multi-embedding approach when keeping the same parameter size.
    This is the most critical result, as it directly validates our primary hypothesis. By simply structuring the embedding space into multiple decoupled representations,  our model gains significant expressive power without increasing the total number of parameters. The inferior performance of the monolithic embedding variant suggests it struggles with feature entanglement and cannot efficiently capture the multifaceted nature of items. This result effectively confirms the existence of the representational bottleneck we hypothesized in the introduction. A single, dense vector, even with the same total dimensionality, lacks the structural inductive bias to effectively disentangle and organize diverse item attributes, leading to a less effective optimization landscape.
    \item The ensemble variant (b) performs worse than our unified Fuxi-MME model while being far less parameter-efficient (it requires $M$ times the number of decoder parameters). This highlights the benefit of shared learning. It also explains that there is a part of sharing information between decomposed representations within the sequential recommendation model. In the ensemble model, each network is isolated and must learn these universal patterns independently, which is less efficient and effective. This proves that enabling input-adaptive parameter specialization within our proposed architecture is a key architectural advantage.
    \item The variant (c) shows a slight decrease in performance compared to our model, indicating that the multi-embedding design is the primary driver of the performance gains. However, the full Fuxi-MME model with MoE still achieves superior results. This demonstrates the synergistic relationship between our two contributions. The multi-embedding layer provides the model with a rich, disentangled set of features, while the MoE layer provides the mechanism for adaptive processing. The MoE-enhanced Fuxi block acts as a dynamic routing network, adaptively sending different feature facets to specialized expert networks for transformation. Without MoE, the model is forced to apply the same dense transformation to all parts of the rich input, which leads to suboptimal performance. The MoE layer provides the final, crucial step of specialization that unlocks the full potential of the multifaceted representations.
\end{itemize}

\subsection{Impact of Hyper-parameter (RQ3)}

To provide practical guidance for deploying Fuxi-MME and to better understand the model's behavior, we conducted a sensitivity analysis on its two most critical hyperparameters: the number of embeddings ($M$) and the number of experts ($n$). All experiments were performed on the Amazon-Beauty dataset, with results presented in Figure \ref{fig:Hyper-parameter}.

\subsubsection{Impact of the Number of Embeddings ($M$)}

To validate the impact of the number of embeddings, we vary the number of embeddings $M$ in the set $\{2, 4, 6, 8\}$ while keeping the total embedding dimension $\mathbf{d}$ fixed. This means the dimension of each sub-embedding ($d/M$) varies, respectively, isolating the effect of representational structure.

As shown in the left subfigure in Figure \ref{fig:Hyper-parameter}, the model's performance exhibits a clear inverted U-shaped trend. Moving from $M=2$ to $M=4$ results in a significant performance improvement, peaking at $\mathbf{M=4}$. This strongly supports our core hypothesis: decomposing the monolithic embedding space allows the model to dedicate different sub-spaces to learning more disentangled and specialized representations of item facets, providing greater representational power. 

Beyond $M=4$, we observe a notable decline in performance. This reveals a critical trade-off. While increasing $M$ promotes feature disentanglement, it simultaneously reduces the dimensionality (capacity) of each individual sub-embedding. When $M$ is large, each sub-embedding has a small dimension, which may be insufficient to capture the complexity of its specialized feature facet. The model suffers from a lack of expressive power within each specialized vector. Therefore, $\mathbf{M=4}$ strikes the optimal balance for this dataset. 

This result highlights a key design consideration: the trade-off between representational diversity (which increases with $\mathbf{M}$) and the representational capacity of each sub-embedding (which decreases with $\mathbf{M}$). For a given total dimension, there exists an optimal point where the model has enough distinct feature channels without making any single channel too simplistic to be useful.

\subsubsection{Impact of the Number of Experts ($n$)}

We vary the number of experts $n$ within the MoE layer from 1 to 8. $n=1$ is equivalent to a dense FFN, serving as a non-MoE baseline. The subfigure in the right of the Figure \ref{fig:Hyper-parameter} illustrates the impact of the number of experts. Increasing the number of experts from 1 to 4 leads to a steady improvement in performance, with the optimum reached at $\mathbf{n=4}$. This demonstrates the value of the MoE architecture: having multiple specialized expert networks allows the model to learn a richer set of transformations and apply the most appropriate one based on the input, handling diverse data patterns more effectively than a single, dense network.

However, a sharp performance drop occurs when $n$ is increased to 6 and beyond, with $n=8$ performing worse than the non-MoE baseline ($n=1$). When we set more experts, the gating network may struggle to assign tokens meaningfully to many experts, leading to underutilized experts and increased noise. Moreover, too many experts dilute the shared knowledge across experts, reducing their collective effectiveness and even underperforming a non-MoE baseline. 


\subsection{An Analysis of MoE Placement (RQ4)}
\begin{table}[t]
\centering
\caption{Model Performance of Different MoE placement.}
\label{tab:attention_ablation}
\renewcommand{\arraystretch}{1.3}
\resizebox{\columnwidth}{!}{
    \begin{tabular}{l|cccc}
    \hhline{=|====}
    \textbf{Model} & \textbf{NG@10} & \textbf{NG@50} & \textbf{HR@10} & \textbf{HR@50}  \\
    \hhline{-|----}
    Fuxi-MME w/o attentionMoE & 0.0344 & 0.0489 & 0.0598 & 0.1262 \\
    w/ qMoE & 0.0326 & 0.0469 & 0.0566 & 0.1224 \\
    w/ kMoE & 0.0332 & 0.0477 & 0.0581 & 0.1252 \\
    w/ vMoE & 0.0305 & 0.0437 & 0.0534 & 0.1140 \\
    \textbf{w/ uMoE} & \textbf{0.0358} & \textbf{0.0508} & \textbf{0.0626} & \textbf{0.1312} \\
    w/o MoE & 0.0331 & 0.0474 & 0.0578 & 0.1232 \\
    \hhline{=|====}
    \end{tabular}
}
\end{table}

The Fuxi block's attention mechanism involves several learnable projection matrices: for queries ($W_q$), keys ($W_k$), values ($W_v$), and the final output gate ($W_u$). While our primary model applies MoE to the Feed-Forward Network and the $W_u$ gate, a natural question arises: is this the optimal placement? To investigate this and better understand the interplay between sparsity and self-attention, we conducted a controlled experiment to determine which projection is most amenable to being replaced by a Mixture-of-Experts layer.

We created several variants of our model. In each variant, we replaced exactly one of the dense projection matrices ($W_q$, $W_k$, $W_v$, or $W_u$) within the attention block with an MoE layer, keeping all other components standard. These variants were then evaluated on the Amazon-Beauty dataset, with the results presented in Table \ref{tab:attention_ablation}.

The variant where the output projection $W_u$ is replaced with an MoE layer is the only one that shows a significant performance improvement. This validates our final architectural choice. The $W_u$ matrix operates on the aggregated, context-aware output of the self-attention mechanism. Placing the MoE layer here allows the model to perform context-dependent transformations. For instance, a specific expert can be activated to process a context vector representing focused brand loyalty, while another handles a context vector representing broad category exploration. This aligns perfectly with our goal of applying specialized, adaptive transformations to the rich outputs generated from our multi-embedding inputs.

In stark contrast, replacing the query, key, or value projection matrices with MoE layers leads to a significant degradation in performance. We attribute this to the fundamental requirements of the self-attention mechanism. The core of self-attention relies on a stable semantic space where queries and keys can be reliably compared. Replacing $W_q$ or $W_k$ with MoE means the projection into this crucial space is handled by different, sparse experts for different tokens. This disrupts the semantic consistency required for meaningful similarity calculations (i.e., Query $\times$ Key). The model needs a single, shared, linear transformation to ensure all items are mapped into a common, comparable space. Besides, replacing $W_v$ with MoE introduces noise and inconsistency into the final aggregated context vector, as the value representations that are weighted and summed are generated by different experts, making it harder for subsequent layers to interpret the output.

In summary, this analysis provides strong evidence that the design of Fuxi-MME preserves the integrity and stability of the self-attention mechanism while introducing powerful, adaptive, and context-aware modulation on its output, thereby validating the design of Fuxi-MME.

\section{Conclusion and Future Work}
In this paper, we addressed a critical yet often overlooked limitation in sequential recommendation: the representational bottleneck imposed by the single embedding layer. While contemporary research has focused on scaling architectural components like attention mechanisms, we argued that the true potential of these models is constrained by the inability of a single vector to capture the multi-faceted nature of items and the lack of adaptive mechanisms to process these features in a context-dependent manner.
To overcome these challenges, we proposed Fuxi-MME, a framework that synergizes a \textbf{multi-embedding strategy} with a \textbf{Mixture-of-Experts (MoE) architecture} upon the powerful Fuxi-$\alpha$ backbone. Specifically, our multi-embedding approach directly tackles the representation issue by decomposing each item’s identity into multiple, lower-dimensional sub-embeddings, enabling a more structured and expressive capture of its diverse attributes. Complementing this, the MoE architecture provides the necessary machinery for dynamic, input-aware processing, routing sequence information to specialized expert networks that can apply the most appropriate transformations based on the context. Our comprehensive experiments on three public benchmarks confirmed the effectiveness of our method. Fuxi-MME not only established a new state-of-the-art by significantly outperforming strong baselines but also demonstrated through ablation studies the critical and complementary roles of both its core components. These results offer a new perspective on the principle of scaling laws in recommendation: instead of relying solely on "brute-force" increases in parameter count, significant performance gains can be unlocked through more principled, structured, and adaptive architectural design.

Looking forward, this work opens several promising avenues for future research. While Fuxi-MME provides a robust framework, the interpretability of its disentangled embeddings and specialized experts remains an exciting direction. Investigating what specific item facets or user behavior patterns are learned by each component could pave the way for more explainable and controllable recommendation systems. Furthermore, the principles of structured representation and adaptive computation can be extended to other domains, such as multi-modal recommendation, where different embedding spaces could naturally correspond to different data modalities (e.g., text, image). By continuing to explore these directions, we believe the architectural concepts pioneered in Fuxi-MME will contribute to building the next generation of more intelligent, expressive, and effective sequential recommendation systems.

\printbibliography

@inproceedings{kang2018self,
  title={Self-attentive sequential recommendation},
  author={Kang, Wang-Cheng and McAuley, Julian},
  booktitle={2018 IEEE international conference on data mining (ICDM)},
  pages={197--206},
  year={2018},
  organization={IEEE}
}

@inproceedings{sun2019bert4rec,
  title={BERT4Rec: Sequential recommendation with bidirectional encoder representations from transformer},
  author={Sun, Fei and Liu, Jun and Wu, Jian and Pei, Changhua and Lin, Xiao and Ou, Wenwu and Jiang, Peng},
  booktitle={Proceedings of the 28th ACM international conference on information and knowledge management},
  pages={1441--1450},
  year={2019}
}

@article{rajput2023recommender,
  title={Recommender systems with generative retrieval},
  author={Rajput, Shashank and Mehta, Nikhil and Singh, Anima and Hulikal Keshavan, Raghunandan and Vu, Trung and Heldt, Lukasz and Hong, Lichan and Tay, Yi and Tran, Vinh and Samost, Jonah and others},
  journal={Advances in Neural Information Processing Systems},
  volume={36},
  pages={10299--10315},
  year={2023}
}

@inproceedings{lee2022autoregressive,
  title={Autoregressive image generation using residual quantization},
  author={Lee, Doyup and Kim, Chiheon and Kim, Saehoon and Cho, Minsu and Han, Wook-Shin},
  booktitle={Proceedings of the IEEE/CVF conference on computer vision and pattern recognition},
  pages={11523--11532},
  year={2022}
}

@article{zhai2024actions,
  title={Actions speak louder than words: Trillion-parameter sequential transducers for generative recommendations},
  author={Zhai, Jiaqi and Liao, Lucy and Liu, Xing and Wang, Yueming and Li, Rui and Cao, Xuan and Gao, Leon and Gong, Zhaojie and Gu, Fangda and He, Michael and others},
  journal={arXiv preprint arXiv:2402.17152},
  year={2024}
}

@inproceedings{zhang2025killing,
  title={Killing two birds with one stone: Unifying retrieval and ranking with a single generative recommendation model},
  author={Zhang, Luankang and Song, Kenan and Lee, Yi Quan and Guo, Wei and Wang, Hao and Li, Yawen and Guo, Huifeng and Liu, Yong and Lian, Defu and Chen, Enhong},
  booktitle={Proceedings of the 48th International ACM SIGIR Conference on Research and Development in Information Retrieval},
  pages={2224--2234},
  year={2025}
}

@article{boka2024survey,
  title={A survey of sequential recommendation systems: Techniques, evaluation, and future directions},
  author={Boka, Tesfaye Fenta and Niu, Zhendong and Neupane, Rama Bastola},
  journal={Information Systems},
  volume={125},
  pages={102427},
  year={2024},
  publisher={Elsevier}
}

@article{hidasi2015session,
  title={Session-based recommendations with recurrent neural networks},
  author={Hidasi, Bal{\'a}zs and Karatzoglou, Alexandros and Baltrunas, Linas and Tikk, Domonkos},
  journal={arXiv preprint arXiv:1511.06939},
  year={2015}
}

@article{xie2024breaking,
  title={Breaking determinism: Fuzzy modeling of sequential recommendation using discrete state space diffusion model},
  author={Xie, Wenjia and Wang, Hao and Zhang, Luankang and Zhou, Rui and Lian, Defu and Chen, Enhong},
  journal={Advances in Neural Information Processing Systems},
  volume={37},
  pages={22720--22744},
  year={2024}
}

@article{li2023ctrl,
  title={Ctrl: Connect tabular and language model for ctr prediction},
  author={Li, Xiangyang and Chen, Bo and Hou, Lu and Tang, Ruiming},
  journal={CoRR},
  year={2023}
}

@article{wu2024survey,
  title={A survey on large language models for recommendation},
  author={Wu, Likang and Zheng, Zhi and Qiu, Zhaopeng and Wang, Hao and Gu, Hongchao and Shen, Tingjia and Qin, Chuan and Zhu, Chen and Zhu, Hengshu and Liu, Qi and others},
  journal={World Wide Web},
  volume={27},
  number={5},
  pages={60},
  year={2024},
  publisher={Springer}
}

@article{shen2023towards,
  title={Towards understanding and mitigating unintended biases in language model-driven conversational recommendation},
  author={Shen, Tianshu and Li, Jiaru and Bouadjenek, Mohamed Reda and Mai, Zheda and Sanner, Scott},
  journal={Information Processing \& Management},
  volume={60},
  number={1},
  pages={103139},
  year={2023},
  publisher={Elsevier}
}

@inproceedings{wang2025generative,
  title={Generative large recommendation models: Emerging trends in llms for recommendation},
  author={Wang, Hao and Guo, Wei and Zhang, Luankang and Chin, Jin Yao and Ye, Yufei and Guo, Huifeng and Liu, Yong and Lian, Defu and Tang, Ruiming and Chen, Enhong},
  booktitle={Companion Proceedings of the ACM on Web Conference 2025},
  pages={49--52},
  year={2025}
}

@article{rafailov2023direct,
  title={Direct preference optimization: Your language model is secretly a reward model},
  author={Rafailov, Rafael and Sharma, Archit and Mitchell, Eric and Manning, Christopher D and Ermon, Stefano and Finn, Chelsea},
  journal={Advances in neural information processing systems},
  volume={36},
  pages={53728--53741},
  year={2023}
}

@article{guo2024scaling,
  title={Scaling new frontiers: Insights into large recommendation models},
  author={Guo, Wei and Wang, Hao and Zhang, Luankang and Chin, Jin Yao and Liu, Zhongzhou and Cheng, Kai and Pan, Qiushi and Lee, Yi Quan and Xue, Wanqi and Shen, Tingjia and others},
  journal={arXiv preprint arXiv:2412.00714},
  year={2024}
}

@inproceedings{liu2024multi,
  title={Multi-behavior generative recommendation},
  author={Liu, Zihan and Hou, Yupeng and McAuley, Julian},
  booktitle={Proceedings of the 33rd ACM International Conference on Information and Knowledge Management},
  pages={1575--1585},
  year={2024}
}

@article{deldjoo2024recommendation,
  title={Recommendation with generative models},
  author={Deldjoo, Yashar and He, Zhankui and McAuley, Julian and Korikov, Anton and Sanner, Scott and Ramisa, Arnau and Vidal, Rene and Sathiamoorthy, Maheswaran and Kasrizadeh, Atoosa and Milano, Silvia and others},
  journal={arXiv preprint arXiv:2409.15173},
  year={2024}
}

@inproceedings{hou2025generative,
  title={Generative Recommendation Models: Progress and Directions},
  author={Hou, Yupeng and Zhang, An and Sheng, Leheng and Yang, Zhengyi and Wang, Xiang and Chua, Tat-Seng and McAuley, Julian},
  booktitle={Companion Proceedings of the ACM on Web Conference 2025},
  pages={13--16},
  year={2025}
}

@article{dai2024deepseekmoe,
  title={Deepseekmoe: Towards ultimate expert specialization in mixture-of-experts language models},
  author={Dai, Damai and Deng, Chengqi and Zhao, Chenggang and Xu, RX and Gao, Huazuo and Chen, Deli and Li, Jiashi and Zeng, Wangding and Yu, Xingkai and Wu, Yu and others},
  journal={arXiv preprint arXiv:2401.06066},
  year={2024}
}

@inproceedings{zoph2022designing,
  title={Designing effective sparse expert models},
  author={Zoph, Barret},
  booktitle={2022 IEEE International Parallel and Distributed Processing Symposium Workshops (IPDPSW)},
  pages={1044--1044},
  year={2022},
  organization={IEEE}
}

@article{li2023exploring,
  title={Exploring the upper limits of text-based collaborative filtering using large language models: Discoveries and insights},
  author={Li, Ruyu and Deng, Wenhao and Cheng, Yu and Yuan, Zheng and Zhang, Jiaqi and Yuan, Fajie},
  journal={arXiv preprint arXiv:2305.11700},
  year={2023}
}

@inproceedings{wang2021siamese,
  title={Siamese generative adversarial predicting network for extremely sparse data in recommendation system},
  author={Wang, Qingxian and Zhang, Renjian and Ma, Kangkang and Chen, Bo and Chen, Jiufang and Shi, Xiaoyu},
  booktitle={2021 IEEE Intl Conf on Parallel \& Distributed Processing with Applications, Big Data \& Cloud Computing, Sustainable Computing \& Communications, Social Computing \& Networking (ISPA/BDCloud/SocialCom/SustainCom)},
  pages={1236--1241},
  year={2021},
  organization={IEEE}
}

@inproceedings{chen2024collaborative,
  title={Collaborative Filtering Algorithm Based on Generative Adversarial Networks},
  author={Chen, Junshu and Liu, GuangCong},
  booktitle={2024 IEEE 14th International Conference on Electronics Information and Emergency Communication (ICEIEC)},
  pages={1--6},
  year={2024},
  organization={IEEE}
}

@inproceedings{zhou2023vcgan,
  title={VCGAN: Variational Collaborative Generative Adversarial Network for Recommendation Systems},
  author={Zhou, Chung-Han and Chen, Yi-Ling},
  booktitle={ICC 2023-IEEE International Conference on Communications},
  pages={6324--6330},
  year={2023},
  organization={IEEE}
}

@inproceedings{wu2024diffusion,
  title={A diffusion data enhancement retentive model for sequential recommendation},
  author={Wu, Tengqing},
  booktitle={2024 7th International Conference on Computer Information Science and Application Technology (CISAT)},
  pages={114--118},
  year={2024},
  organization={IEEE}
}

@article{chen2022double,
  title={Double attention convolutional neural network for sequential recommendation},
  author={Chen, Qi and Li, Guohui and Zhou, Quan and Shi, Si and Zou, Deqing},
  journal={ACM Transactions on the Web},
  volume={16},
  number={4},
  pages={1--23},
  year={2022},
  publisher={ACM New York, NY}
}

@inproceedings{hidasi2018recurrent,
  title={Recurrent neural networks with top-k gains for session-based recommendations},
  author={Hidasi, Bal{\'a}zs and Karatzoglou, Alexandros},
  booktitle={Proceedings of the 27th ACM international conference on information and knowledge management},
  pages={843--852},
  year={2018}
}

@inproceedings{yan2019cosrec,
  title={CosRec: 2D convolutional neural networks for sequential recommendation},
  author={Yan, An and Cheng, Shuo and Kang, Wang-Cheng and Wan, Mengting and McAuley, Julian},
  booktitle={Proceedings of the 28th ACM international conference on information and knowledge management},
  pages={2173--2176},
  year={2019}
}

@inproceedings{xu2019survey,
  title={A survey on sequential recommendation},
  author={Xu, Mingming and Liu, Fangai and Xu, Weizhi},
  booktitle={2019 6th international conference on information science and control engineering (ICISCE)},
  pages={106--111},
  year={2019},
  organization={IEEE}
}

@article{pan2024survey,
  title={A Survey on Sequential Recommendation},
  author={Pan, Liwei and Pan, Weike and Wei, Meiyan and Yin, Hongzhi and Ming, Zhong},
  journal={arXiv preprint arXiv:2412.12770},
  year={2024}
}

@article{deng2025onerec,
  title={Onerec: Unifying retrieve and rank with generative recommender and iterative preference alignment},
  author={Deng, Jiaxin and Wang, Shiyao and Cai, Kuo and Ren, Lejian and Hu, Qigen and Ding, Weifeng and Luo, Qiang and Zhou, Guorui},
  journal={arXiv preprint arXiv:2502.18965},
  year={2025}
}

@inproceedings{he2016fusing,
  title={Fusing similarity models with markov chains for sparse sequential recommendation},
  author={He, Ruining and McAuley, Julian},
  booktitle={2016 IEEE 16th international conference on data mining (ICDM)},
  pages={191--200},
  year={2016},
  organization={IEEE}
}

@article{zhang2024wukong,
  title={Wukong: Towards a scaling law for large-scale recommendation},
  author={Zhang, Buyun and Luo, Liang and Chen, Yuxin and Nie, Jade and Liu, Xi and Guo, Daifeng and Zhao, Yanli and Li, Shen and Hao, Yuchen and Yao, Yantao and others},
  journal={arXiv preprint arXiv:2403.02545},
  year={2024}
}

@article{guo2023embedding,
  title={On the embedding collapse when scaling up recommendation models},
  author={Guo, Xingzhuo and Pan, Junwei and Wang, Ximei and Chen, Baixu and Jiang, Jie and Long, Mingsheng},
  journal={arXiv preprint arXiv:2310.04400},
  year={2023}
}

@article{guo2025onesug,
  title={OneSug: The Unified End-to-End Generative Framework for E-commerce Query Suggestion},
  author={Guo, Xian and Chen, Ben and Wang, Siyuan and Yang, Ying and Lei, Chenyi and Ding, Yuqing and Li, Han},
  journal={arXiv preprint arXiv:2506.06913},
  year={2025}
}

@inproceedings{rendle2010factorizing,
  title={Factorizing personalized markov chains for next-basket recommendation},
  author={Rendle, Steffen and Freudenthaler, Christoph and Schmidt-Thieme, Lars},
  booktitle={Proceedings of the 19th international conference on World wide web},
  pages={811--820},
  year={2010}
}

@inproceedings{zhou2020s3,
  title={S3-rec: Self-supervised learning for sequential recommendation with mutual information maximization},
  author={Zhou, Kun and Wang, Hui and Zhao, Wayne Xin and Zhu, Yutao and Wang, Sirui and Zhang, Fuzheng and Wang, Zhongyuan and Wen, Ji-Rong},
  booktitle={Proceedings of the 29th ACM international conference on information \& knowledge management},
  pages={1893--1902},
  year={2020}
}

@inproceedings{he2014practical,
  title={Practical lessons from predicting clicks on ads at facebook},
  author={He, Xinran and Pan, Junfeng and Jin, Ou and Xu, Tianbing and Liu, Bo and Xu, Tao and Shi, Yanxin and Atallah, Antoine and Herbrich, Ralf and Bowers, Stuart and others},
  booktitle={Proceedings of the eighth international workshop on data mining for online advertising},
  pages={1--9},
  year={2014}
}

@article{touvron2023llama,
  title={Llama: Open and efficient foundation language models},
  author={Touvron, Hugo and Lavril, Thibaut and Izacard, Gautier and Martinet, Xavier and Lachaux, Marie-Anne and Lacroix, Timoth{\'e}e and Rozi{\`e}re, Baptiste and Goyal, Naman and Hambro, Eric and Azhar, Faisal and others},
  journal={arXiv preprint arXiv:2302.13971},
  year={2023}
}

@article{yang2025qwen3,
  title={Qwen3 technical report},
  author={Yang, An and Li, Anfeng and Yang, Baosong and Zhang, Beichen and Hui, Binyuan and Zheng, Bo and Yu, Bowen and Gao, Chang and Huang, Chengen and Lv, Chenxu and others},
  journal={arXiv preprint arXiv:2505.09388},
  year={2025}
}

@article{medsker2001recurrent,
  title={Recurrent neural networks},
  author={Medsker, Larry R and Jain, Lakhmi and others},
  journal={Design and applications},
  volume={5},
  number={64-67},
  pages={2},
  year={2001}
}

@article{hochreiter1997long,
  title={Long short-term memory},
  author={Hochreiter, Sepp and Schmidhuber, J{\"u}rgen},
  journal={Neural computation},
  volume={9},
  number={8},
  pages={1735--1780},
  year={1997},
  publisher={MIT press}
}

@article{vaswani2017attention,
  title={Attention is all you need},
  author={Vaswani, Ashish and Shazeer, Noam and Parmar, Niki and Uszkoreit, Jakob and Jones, Llion and Gomez, Aidan N and Kaiser, {\L}ukasz and Polosukhin, Illia},
  journal={Advances in neural information processing systems},
  volume={30},
  year={2017}
}

@article{kaplan2020scaling,
  title={Scaling laws for neural language models},
  author={Kaplan, Jared and McCandlish, Sam and Henighan, Tom and Brown, Tom B and Chess, Benjamin and Child, Rewon and Gray, Scott and Radford, Alec and Wu, Jeffrey and Amodei, Dario},
  journal={arXiv preprint arXiv:2001.08361},
  year={2020}
}

@article{shen2024optimizing,
  title={Optimizing sequential recommendation models with scaling laws and approximate entropy},
  author={Shen, Tingjia and Wang, Hao and Wu, Chuhan and Chin, Jin Yao and Guo, Wei and Liu, Yong and Guo, Huifeng and Lian, Defu and Tang, Ruiming and Chen, Enhong},
  journal={arXiv preprint arXiv:2412.00430},
  year={2024}
}

@inproceedings{ye2025fuxi,
  title={FuXi-$\alpha$: Scaling Recommendation Model with Feature Interaction Enhanced Transformer},
  author={Ye, Yufei and Guo, Wei and Chin, Jin Yao and Wang, Hao and Zhu, Hong and Lin, Xi and Ye, Yuyang and Liu, Yong and Tang, Ruiming and Lian, Defu and others},
  booktitle={Companion Proceedings of the ACM on Web Conference 2025},
  pages={557--566},
  year={2025}
}

@article{wang2025universal,
  title={A universal framework for compressing embeddings in ctr prediction},
  author={Wang, Kefan and Wang, Hao and Song, Kenan and Guo, Wei and Cheng, Kai and Li, Zhi and Liu, Yong and Lian, Defu and Chen, Enhong},
  journal={arXiv preprint arXiv:2502.15355},
  year={2025}
}

@inproceedings{jiang2021xlightfm,
  title={xLightFM: Extremely memory-efficient factorization machine},
  author={Jiang, Gangwei and Wang, Hao and Chen, Jin and Wang, Haoyu and Lian, Defu and Chen, Enhong},
  booktitle={Proceedings of the 44th International ACM SIGIR Conference on Research and Development in Information Retrieval},
  pages={337--346},
  year={2021}
}

@article{zhang2024gme,
  title={GME: Improving Universal Multimodal Retrieval by Multimodal LLMs},
  author={Zhang, Xin and Zhang, Yanzhao and Xie, Wen and Li, Mingxin and Dai, Ziqi and Long, Dingkun and Xie, Pengjun and Zhang, Meishan and Li, Wenjie and Zhang, Min},
  journal={arXiv preprint arXiv:2412.16855},
  year={2024}
}

@inproceedings{li2017neural,
  title={Neural attentive session-based recommendation},
  author={Li, Jing and Ren, Pengjie and Chen, Zhumin and Ren, Zhaochun and Lian, Tao and Ma, Jun},
  booktitle={Proceedings of the 2017 ACM on Conference on Information and Knowledge Management},
  pages={1419--1428},
  year={2017}
}

@article{fan2025tim4rec,
  title={Tim4rec: An efficient sequential recommendation model based on time-aware structured state space duality model},
  author={Fan, Hao and Zhu, Mengyi and Hu, Yanrong and Feng, Hailin and He, Zhijie and Liu, Hongjiu and Liu, Qingyang},
  journal={Neurocomputing},
  pages={131270},
  year={2025},
  publisher={Elsevier}
}

@article{liu2024mamba4rec,
  title={Mamba4rec: Towards efficient sequential recommendation with selective state space models},
  author={Liu, Chengkai and Lin, Jianghao and Wang, Jianling and Liu, Hanzhou and Caverlee, James},
  journal={arXiv preprint arXiv:2403.03900},
  year={2024}
}

@inproceedings{klenitskiy2023turning,
  title={Turning dross into gold loss: is bert4rec really better than sasrec?},
  author={Klenitskiy, Anton and Vasilev, Alexey},
  booktitle={Proceedings of the 17th ACM Conference on Recommender Systems},
  pages={1120--1125},
  year={2023}
}

@inproceedings{xie2022contrastive,
  title={Contrastive learning for sequential recommendation},
  author={Xie, Xu and Sun, Fei and Liu, Zhaoyang and Wu, Shiwen and Gao, Jinyang and Zhang, Jiandong and Ding, Bolin and Cui, Bin},
  booktitle={2022 IEEE 38th international conference on data engineering (ICDE)},
  pages={1259--1273},
  year={2022},
  organization={IEEE}
}

@inproceedings{yin2024dataset,
  title={Dataset regeneration for sequential recommendation},
  author={Yin, Mingjia and Wang, Hao and Guo, Wei and Liu, Yong and Zhang, Suojuan and Zhao, Sirui and Lian, Defu and Chen, Enhong},
  booktitle={Proceedings of the 30th ACM SIGKDD Conference on Knowledge Discovery and Data Mining},
  pages={3954--3965},
  year={2024}
}

@inproceedings{chen2022intent,
  title={Intent contrastive learning for sequential recommendation},
  author={Chen, Yongjun and Liu, Zhiwei and Li, Jia and McAuley, Julian and Xiong, Caiming},
  booktitle={Proceedings of the ACM web conference 2022},
  pages={2172--2182},
  year={2022}
}

@article{kingma2014adam,
  title={Adam: A method for stochastic optimization},
  author={Kingma, Diederik P and Ba, Jimmy},
  journal={arXiv preprint arXiv:1412.6980},
  year={2014}
}

@inproceedings{liu2023user,
  title={User Behavior Modeling with Deep Learning for Recommendation: Recent Advances},
  author={Liu, Weiwen and Guo, Wei and Liu, Yong and Tang, Ruiming and Wang, Hao},
  booktitle={Proceedings of the 17th ACM Conference on Recommender Systems},
  pages={1286--1287},
  year={2023}
}

@misc{shen2025optimizingsequentialrecommendationmodels,
      title={Optimizing Sequential Recommendation Models with Scaling Laws and Approximate Entropy}, 
      author={Tingjia Shen and Hao Wang and Chuhan Wu and Jin Yao Chin and Wei Guo and Yong Liu and Huifeng Guo and Defu Lian and Ruiming Tang and Enhong Chen},
      year={2025},
      eprint={2412.00430},
      archivePrefix={arXiv},
      primaryClass={cs.AI},
      url={https://arxiv.org/abs/2412.00430}, 
}

@article{zhang2025td3,
  title={TD3: Tucker Decomposition Based Dataset Distillation Method for Sequential Recommendation},
  author={Zhang, Jiaqing and Yin, Mingjia and Wang, Hao and Li, Yawen and Ye, Yuyang and Lou, Xingyu and Du, Junping and Chen, Enhong},
  journal={arXiv preprint arXiv:2502.02854},
  year={2025}
}

@article{wang2025mf,
  title={MF-GSLAE: A Multi-Factor User Representation Pre-training Framework for Dual-Target Cross-Domain Recommendation},
  author={Wang, Hao and Yin, Mingjia and Zhang, Luankang and Zhao, Sirui and Chen, Enhong},
  journal={ACM Transactions on Information Systems},
  volume={43},
  number={2},
  pages={1--28},
  year={2025},
  publisher={ACM New York, NY}
}

@inproceedings{wang2019mcne,
  title={MCNE: An end-to-end framework for learning multiple conditional network representations of social network},
  author={Wang, Hao and Xu, Tong and Liu, Qi and Lian, Defu and Chen, Enhong and Du, Dongfang and Wu, Han and Su, Wen},
  booktitle={Proceedings of the 25th ACM SIGKDD international conference on knowledge discovery \& data mining},
  pages={1064--1072},
  year={2019}
}

@article{wang2021hypersorec,
  title={Hypersorec: Exploiting hyperbolic user and item representations with multiple aspects for social-aware recommendation},
  author={Wang, Hao and Lian, Defu and Tong, Hanghang and Liu, Qi and Huang, Zhenya and Chen, Enhong},
  journal={ACM Transactions on Information Systems (TOIS)},
  volume={40},
  number={2},
  pages={1--28},
  year={2021},
  publisher={ACM New York, NY}
}

@article{wang2021decoupled,
  title={Decoupled representation learning for attributed networks},
  author={Wang, Hao and Lian, Defu and Tong, Hanghang and Liu, Qi and Huang, Zhenya and Chen, Enhong},
  journal={IEEE Transactions on Knowledge and Data Engineering},
  volume={35},
  number={3},
  pages={2430--2444},
  year={2021},
  publisher={IEEE}
}

@inproceedings{tong2024mdap,
  title={MDAP: A Multi-view Disentangled and Adaptive Preference Learning Framework for Cross-Domain Recommendation},
  author={Tong, Junxiong and Yin, Mingjia and Wang, Hao and Pan, Qiushi and Lian, Defu and Chen, Enhong},
  booktitle={International Conference on Web Information Systems Engineering},
  pages={164--178},
  year={2024},
  organization={Springer}
}

@inproceedings{zhang2024unified,
  title={A Unified Framework for Adaptive Representation Enhancement and Inversed Learning in Cross-Domain Recommendation},
  author={Zhang, Luankang and Wang, Hao and Zhang, Suojuan and Yin, Mingjia and Han, Yongqiang and Zhang, Jiaqing and Lian, Defu and Chen, Enhong},
  booktitle={International Conference on Database Systems for Advanced Applications},
  pages={115--130},
  year={2024},
  organization={Springer}
}

@article{yin2024learning,
  title={Learning partially aligned item representation for cross-domain sequential recommendation},
  author={Yin, Mingjia and Wang, Hao and Guo, Wei and Liu, Yong and Li, Zhi and Zhao, Sirui and Wang, Zhen and Lian, Defu and Chen, Enhong},
  journal={arXiv preprint arXiv:2405.12473},
  year={2024}
}

@article{ye2025fuxi2,
  title={FuXi-$\backslash$beta: Towards a Lightweight and Fast Large-Scale Generative Recommendation Model},
  author={Ye, Yufei and Guo, Wei and Wang, Hao and Zhu, Hong and Ye, Yuyang and Liu, Yong and Guo, Huifeng and Tang, Ruiming and Lian, Defu and Chen, Enhong},
  journal={arXiv preprint arXiv:2508.10615},
  year={2025}
}

@article{xu2024multi,
  title={Multi-granularity Interest Retrieval and Refinement Network for Long-Term User Behavior Modeling in CTR Prediction},
  author={Xu, Xiang and Wang, Hao and Guo, Wei and Zhang, Luankang and Yang, Wanshan and Yu, Runlong and Liu, Yong and Lian, Defu and Chen, Enhong},
  journal={arXiv preprint arXiv:2411.15005},
  year={2024}
}

@inproceedings{yin2023apgl4sr,
  title={Apgl4sr: A generic framework with adaptive and personalized global collaborative information in sequential recommendation},
  author={Yin, Mingjia and Wang, Hao and Xu, Xiang and Wu, Likang and Zhao, Sirui and Guo, Wei and Liu, Yong and Tang, Ruiming and Lian, Defu and Chen, Enhong},
  booktitle={Proceedings of the 32nd ACM international conference on information and knowledge management},
  pages={3009--3019},
  year={2023}
}

@article{wang2024denoising,
  title={Denoising Pre-Training and Customized Prompt Learning for Efficient Multi-Behavior Sequential Recommendation},
  author={Wang, Hao and Han, Yongqiang and Wang, Kefan and Cheng, Kai and Wang, Zhen and Guo, Wei and Liu, Yong and Lian, Defu and Chen, Enhong},
  journal={arXiv preprint arXiv:2408.11372},
  year={2024}
}

@inproceedings{han2024efficient,
  title={Efficient noise-decoupling for multi-behavior sequential recommendation},
  author={Han, Yongqiang and Wang, Hao and Wang, Kefan and Wu, Likang and Li, Zhi and Guo, Wei and Liu, Yong and Lian, Defu and Chen, Enhong},
  booktitle={Proceedings of the ACM Web Conference 2024},
  pages={3297--3306},
  year={2024}
}

@inproceedings{han2023guesr,
  title={Guesr: A global unsupervised data-enhancement with bucket-cluster sampling for sequential recommendation},
  author={Han, Yongqiang and Wu, Likang and Wang, Hao and Wang, Guifeng and Zhang, Mengdi and Li, Zhi and Lian, Defu and Chen, Enhong},
  booktitle={International conference on database systems for advanced applications},
  pages={286--296},
  year={2023},
  organization={Springer}
}

@article{gu2025rapid,
  title={RAPID: Efficient Retrieval-Augmented Long Text Generation with Writing Planning and Information Discovery},
  author={Gu, Hongchao and Li, Dexun and Dong, Kuicai and Zhang, Hao and Lv, Hang and Wang, Hao and Lian, Defu and Liu, Yong and Chen, Enhong},
  journal={arXiv preprint arXiv:2503.00751},
  year={2025}
}

@article{yin2024entropy,
  title={Entropy law: The story behind data compression and llm performance},
  author={Yin, Mingjia and Wu, Chuhan and Wang, Yufei and Wang, Hao and Guo, Wei and Wang, Yasheng and Liu, Yong and Tang, Ruiming and Lian, Defu and Chen, Enhong},
  journal={arXiv preprint arXiv:2407.06645},
  year={2024}
}

@article{shen2024exploring,
  title={Exploring user retrieval integration towards large language models for cross-domain sequential recommendation},
  author={Shen, Tingjia and Wang, Hao and Zhang, Jiaqing and Zhao, Sirui and Li, Liangyue and Chen, Zulong and Lian, Defu and Chen, Enhong},
  journal={arXiv preprint arXiv:2406.03085},
  year={2024}
}

@misc{shen2025genkienhancingopendomainquestion,
      title={GenKI: Enhancing Open-Domain Question Answering with Knowledge Integration and Controllable Generation in Large Language Models}, 
      author={Tingjia Shen and Hao Wang and Chuan Qin and Ruijun Sun and Yang Song and Defu Lian and Hengshu Zhu and Enhong Chen},
      year={2025},
      eprint={2505.19660},
      archivePrefix={arXiv},
      primaryClass={cs.CL},
      url={https://arxiv.org/abs/2505.19660}, 
}

@misc{lv2025costeer,
    title={CoSteer: Collaborative Decoding-Time Personalization via Local Delta Steering},
    author={Hang Lv and Sheng Liang and Hao Wang and Hongchao Gu and Yaxiong Wu and Wei Guo and Defu Lian and Yong Liu and Enhong Chen},
    year={2025},
    eprint={2507.04756},
    archivePrefix={arXiv},
    primaryClass={cs.CL}
}

@article{huang2024chemeval,
  title={ChemEval: A Comprehensive Multi-Level Chemical Evaluation for Large Language Models},
  author={Huang, Yuqing and Zhang, Rongyang and He, Xuesong and Zhi, Xuyang and Wang, Hao and Li, Xin and Xu, Feiyang and Liu, Deguang and Liang, Huadong and Li, Yi and others},
  journal={arXiv preprint arXiv:2409.13989},
  year={2024}
}

\end{document}